\input{aipcheck}

\documentclass[
    ,final            
  ]
  {aipproc}

\layoutstyle{6x9}

\begin{document}

\title{Coordinates and frames from the causal\\ point of
view}

\classification{04.20.-q} \keywords {space-time frames, causal
structure}

\author{Juan Antonio Morales Lladosa}{
  address={Departament
d'Astronomia i Astrof\'{\i}sica, Universitat de Val\`encia, 46100
Burjassot, Val\`encia, Spain}}

\begin{abstract}
Lorentzian frames may belong to one of the $199$ causal classes.
Of these numerous causal classes, people are essentially aware
only of two of them. Nevertheless, other causal classes are
present in some well-known solutions, or present a strong interest
in the physical construction of coordinate systems. Here we show
the unusual causal classes to which belong so familiar coordinate
systems as those of Lemaître, those of Eddington-Finkelstein, or
those of Bondi-Sachs. Also the causal classes associated to the
Coll light coordinates (four congruences of real geodetic null
lines) and to the Coll positioning systems (light signals
broadcasted by four clocks) are analyzed. The role that these
results play in the comprehension and classification of
relativistic coordinate systems is emphasized.
\end{abstract}

\maketitle


The main purpose of this short communication is to gain stimulus
in the investigation of relativistic coordinate systems from the
causal point of view. This issue might be relevant in several
situations. For example, to investigate the coordinates which are
appropriated to deal with evolution problems throughout horizons.
Current $3+1$ numerical codes in relativistic hydrodynamics are
being implemented using such a coordinates. Also, to consider
admissible {\it cuts} of the space-time others than the very usual
``space $\oplus$ time'' decomposition. The causal classification
of frames may help us to better understand other aspects of the
space-time, for instance the "light $\oplus$ light $\oplus$ light
$\oplus$ light" decomposition.

Firstly, we remember the causal classification of Lorentzian
frames (also presented at the ERE-88 celebrated in Salamanca).
Then, we give several examples of causal classes of relativistic
coordinates: those associated with the coordinates introduced by
Lemaître, by Bondi and Sachs, and by Coll.

In dimension $n=4$, the causal class of a frame $\{\rm{v_1, v_2,
v_3, v_4}\}$ is defined by a set of 14 characters:
$$
\{{\rm c_1 c_2 c_3 c_4}, {\rm C_{12}C_{13} C_{14} C_{23} C_{24} C_{34}}, {\it c_1 c_2 c_3 c_4}\}
$$
${\rm c_i}$ being the causal character of the vector ${\rm v_i}$,
${\rm C_{ij}}$ (${\rm i \neq j}$) being the causal character of
the adjoint 2-plane $\{{\rm v_i v_j}\}$, and ${\it c_i}$ being the
causal character of the covectors of the dual coframe
$\{\theta^{\rm i}\}$, $\theta^{\rm i} (\rm{v}_j) = \delta^{\rm i}
_{\rm \, j}$. The covector $\theta^{\rm i}$ is time-like (resp.
space-like) iff the $3$-plane generated by $\{\rm v_j\}_{\rm  {j
\neq i}}$ is space-like (resp. space-like). This applies  for both
the Newtonian and the Lorentzian causal structures. In addition,
for the later, the covector $\theta^{\rm i}$ is light-like iff the
$3$-plane generated by $\{\rm v_j\}_{\rm{j \neq i}}$ is
light-like. Elsewhere (see \cite{Salamanca} and \cite{199}) we
have presented the following result:\\

\underline{Theorem:} {\em i) In the 4-dimensional Newtonian
space-time there exist 4, and only 4, causal classes of frames,
and ii) In the 4-dimensional relativistic space-time there exist
199, and only 199, causal classes of Lorentzian frames.}

\vspace{1mm}

This result provides the causal classification of coordinate
systems in Newtonian and in Relativistic physics. A coordinate
system $\{x^1, x^2, x^3, x^4\}$ belongs to the causal class
$\{{\rm c_1 c_2 c_3 c_4}, {\rm C_{12} C_{13} C_{14} C_{23} C_{24}
C_{34}}, {\it c_1 c_2 c_3 c_4}\}$ if the cobasis $\{d x^1, d x^2,
d x^3, d x^4\}$ has causal type $({\it c_1 c_2 c_3 c_4})$ and has
associated four families of coordinate 3-surfaces whose mutual
intersections give six families of coordinate 2-surfaces of causal
characters $({\rm C_{12} C_{13} C_{14} C_{23} C_{24} C_{34}})$ and
four congruences of coordinate lines of causal characters $({\rm
c_1 c_2 c_3 c_4})$.

As a matter of notation, romanic letters ($\rm{e, t, l}$)
represent the causal character of vectors (space-like, time-like,
light-like , respectively), and capital ($\rm{E, T , L}$) and
italic letters ($\it{e, t, l}$) denote the causal character of
2-planes and covectors, respectively. Accordingly, the four
Newtonian causal classes are denoted as:
$$\{{\rm t e e e, T T T E E E}, {\it
t e e e}\}, \{{\rm t t e e, T T T T T E}, {\it e e e e}\}, \{{\rm
t t t e, T T T T T T}, {\it e e e e}\}, \{{\rm t t t t, T T T T T
T}, {\it e e e e}\}.$$
Among the 199 Lorentzian classes, four of them have the same set
of causal characters as the Newtonian ones \cite{Salamanca}. Next,
we consider another examples of relativistic classes.

\vspace{1mm}

{\it 1.- The three causal classes of Lemaître coordinates.} The
familiar metric form of the Schwarzschild solution in coordinates
$\{t, r, \theta,
 \phi\}$ is written as
$$
 ds^2=- \left(1-\frac{2m}{r}\right)\, dt^2 + \frac {1}{\displaystyle{1- \frac
{2m}{r}}} \,\, d r^2 + r^2 d\Omega^2
$$
where $r > 2m$ and $d\Omega^2 = d\theta^2 + \sin^2 \theta \,
d\phi^2$. The coordinate basis $\{\partial_t, \partial_r,
\partial_\theta,
\partial_\phi\}$ belong to the causal class
$\{{\rm t e e e, T T T E E E}, {\it t e e e}\}$. Lema{\^{\i}}tre
\cite{Elmestre} extended the Schwarzschild solution at the region
$0<r\leq 2m$, obtaining a metric form
$$
 ds^2=- \left(1-\frac{2m}{r}\right)\, dT^2 + 2 \epsilon \sqrt{\frac
{2m}{r}} \, d T \, d r + d r^2 + r^2 d\Omega^2 \qquad
(\epsilon=\pm 1)
$$
that is regular at $r=2m$. The causal character of the coordinate
lines are given by the sign of the four diagonal elements
$g_{\alpha\alpha}$ of the metric $g_{\alpha \beta}$ in this basis.
The causal character of the coordinate 2-surfaces is given by the
sign of the principal second order minors $g_{\alpha \alpha}
g_{\beta \beta} -(g_{\alpha\beta})^2$. And the causal character of
the coordinate 3-surfaces is related to the sign of the diagonal
elements $g^{\alpha\alpha}$ of the contravariant metric expression
$g^{\alpha \beta}$. Consequently, the Lemaître coordinate basis
$\{\partial_T, \partial_r, \partial_\theta, \partial_\phi\}$
belong to the causal class $\{{\rm t e e e, T T T E E E}, {\it t e
e e} \}$ if $r>2m$, $\{{\rm l e e e, T L L E E E}, {\it t l e e}
\}$ if $r=2m$,  or $\{{\rm e e e e, T E E E E  E}, \,{\it t t e e}
\}$ if $r<2m$.

\vspace{1mm}

{\it 2.- The thirteen causal classes of Bondi-Sachs coordinates.}
 Any space-time metric may be expressed in the form \cite{Sachs}:
\begin{displaymath}
g_{\alpha \beta} \equiv g\left(\frac{\partial}{\partial x^\alpha},
\frac{\partial}{\partial x^\beta}\right) = \left(
\begin{array}{cccc}
g_{00} & g_{01} & g_{02}& g_{03} \\
g_{01} & \, 0 & 0 & 0 \\
g_{02} & \, 0 & g_{22} & g_{23} \\
g_{03} & \, 0 & g_{23} & g_{33} \\
\end{array} \right)
\end{displaymath}
For the associated contravariant metric one has
$g^{00}=g^{02}=g^{03}=0$. Therefore,  $\{x^0= \lambda\}$  (with
$\lambda$ a real parameter) are null hypersurfaces. When $g_{22}
g_{33}- g_{23}^2 > 0$, the coordinates are called (generalized)
Bondi-Sachs coordinates \cite{Sachs} \cite{Flet-Lun}, and are
usually denoted by $x^0 \equiv u$, $x^1 \equiv r$, $x^2 \equiv
\theta$, $x^3 \equiv \phi$. We have that {\em the Bondi-Sachs
coordinate systems are classified in 13 causal classes:}

\begin{displaymath}
\begin{array}{ccc}

\{{\rm t \, l \,  e \, e, \,  T  \, T  \, T  \, L  \, L  \, E}, \,
{\it l \, e \,  e \,  e}\}& \qquad  & \\
\\
\{{\rm l \,  l \,  e \,  e, \, T  \, T  \, T  \, L  \, L  \, E},
\, {\it l  \,  e \,  e  \, e}\}& \qquad \{{\rm e  \, l  \, e  \,
e, \, T  \, T  \, T  \, L  \, L  \, E},
{\it l  \, e  \, e  \, e}\} & \\

\{{\rm l  \, l  \, e  \, e, \,  T  \,  T  \, L  \, L  \, L  \, E},
\, {\it l  \, e  \, e  \, e}\} & \qquad \{{\rm e  \, l  \, e  \,
e, \, T  \, T  \, L  \, L  \, L  \, E}, {\it l  \,  e  \,  e  \,
e}\}& \\

\{{\rm l  \, l  \, e  \, e,  \, T  \, L  \, L  \, L \,  L \,  E},
\, {\it l  \,  l  \,  e  \, e}\,\}&  \qquad \{{\rm e  \, l  \, e
\, e,  \, T  \, L  \, L  \, L  \, L  \, E}, \, {\it l  \, e  \, e
\,e}\} & \\
\\
\{{\rm l  \, e  \, e  \, e, T  \, L  \, L  \, L  \, E  \, E},  \,
{\it l \,  l  \, e  \, e}\}& \qquad \{{\rm e  \, l  \, e  \, e,
\, T  \, T  \, E  \, L  \, L  \, E},
{\it l  \, e  \, e  \, e}\}& \\

\{{\rm l  \, e  \, e  \, e,  \, T  \, L  \, L  \, E  \, E  \, E},
\, {\it l  \, l  \, e  \, e}\} & \qquad \{{\rm e  \, l  \, e  \,
e, \, T  \, L  \, E  \, L  \, L  \, E},
{\it l  \, e  \, e  \, e}\} & \\

\{{\rm l  \, e  \, e  \, e,  \, T  \, L  \, L  \, E  \, E  \, E},
\, {\it t  \, l  \, e  \, e}\} & \qquad \{{\rm e  \, l  \, e  \,
e, \, T  \, E  \, E  \, L  \, L  \,  E},
\, {\it l  \, e  \, e  \, e}\}& \\

\end{array}
\end{displaymath}

For example, a coordinate system of the causal
 class $\{{\rm elee,TEELLE}, {\it leee}\}$ has associated a family of
null coordinate 3-surfaces and three families of time-like
3-surfaces. Their mutual cuts give one family of time-like
surfaces, two families of null 2-surfaces and three families of
space-like 2-surfaces. The intersections of these surfaces give a
congruence of null lines and three congruences of space-like
lines.

Note that the Bondi-Sachs coordinates are a generalization of the
familiar Eddington-Finkelstein coordinates used in the
Schwarzschild space-time.  These coordinates belong to the class
$\{{\rm t l e e, T T T L L E}, {\it l e e e}\}$ outside the
horizon, to the class $\{{\rm l l e e, T L L L L E}, {\it l l e
e}\}$ at the horizon $r=2m$, and to the class $\{{\rm e l e e, T E
E L L E}, {\it l t e e}\}$ inside the horizon. The later is the
class $\{{\rm l e e e, T L L E E E}, {\it t l e e}\}$ (as
 it results when the first and the second vectors are changed).

\vspace{1mm}

{\it 3.- The two causal classes of Coll coordinates.} Let us
consider the classes
$$\{{\rm e\,e\,e\,e,\, E\,E\,E\,E\,E\,E}, {\it l\,l\,l\,l}\}
\quad {\rm  and} \quad \{{\rm l\,l\,l\,l,\, T\,T\,T\,T\,T\,T},
{\it e\,e\,e\,e}\}$$ A coordinate systems of the first class has
associated four families of null 3-surfaces whose mutual cuts give
six families of space-like 2-surfaces and four congruences of
space-like lines. This class includes the {\it emission
coordinates} of the {\it Coll positioning systems}
\cite{Tolo-ERE05} \cite{Joan-ERE05}. A coordinate system of the
second class has associated four families of time-like 3-surfaces
whose mutual cuts give six families of time-like 2-surfaces and
four congruences of null lines.  This class includes the {\it Coll
light coordinates} built from the intersection of four beams of
light \cite{c-luz}. We call them the {\it Coll causal classes}.\\

The 199 causal classes of relativistic coordinates may be wholly
visualized in a table that we call the 199-Table (for details, see
\cite{Salamanca} and \cite{199}). Here, we reproduce the 199-Table
showing the location of the Coll causal classes.

\begin{figure}
\includegraphics[height=1.0\textheight]{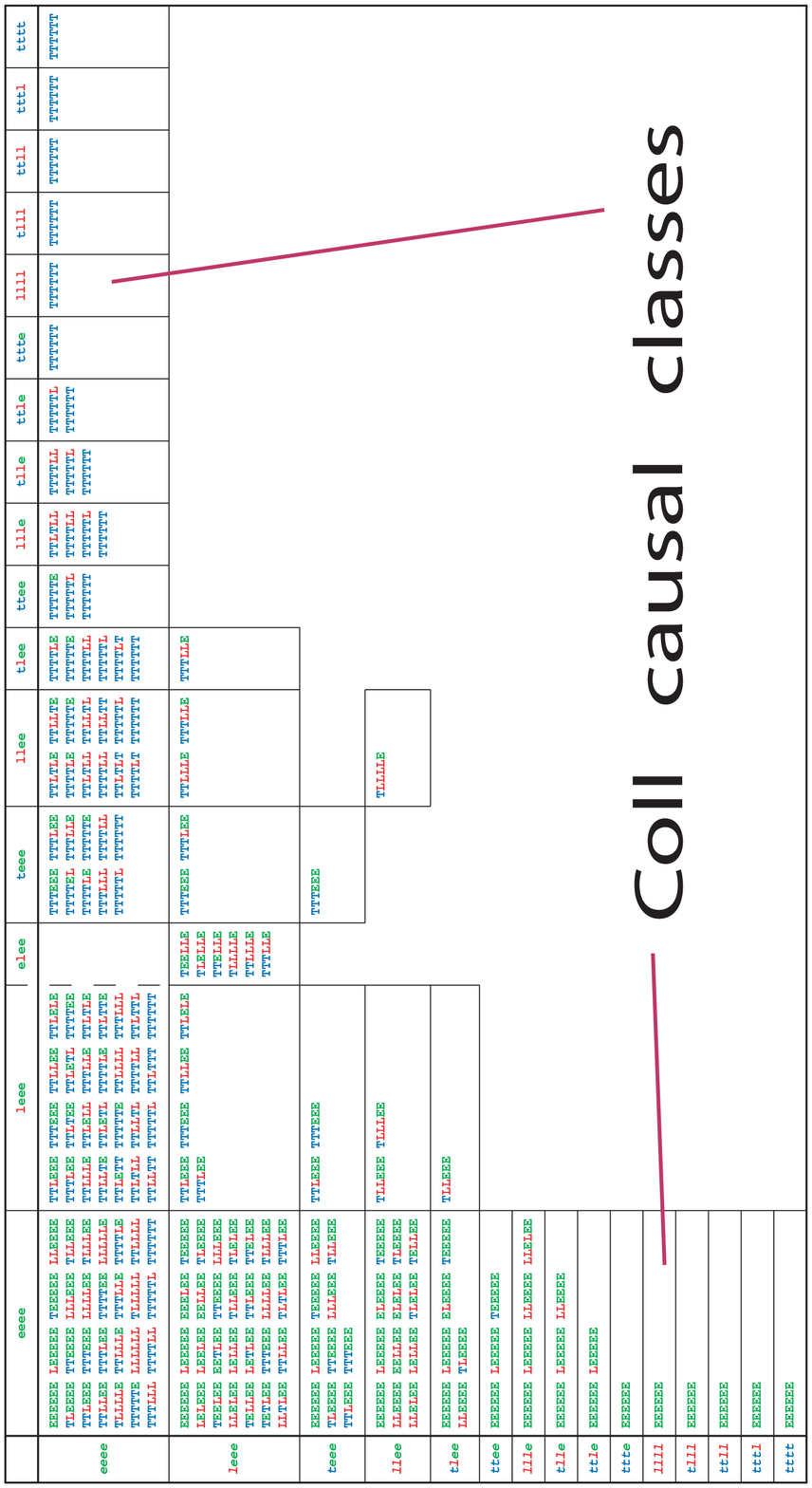}
\caption{The Coll causal classes $\{{\rm l l l l, T T T T T T}\,
{\it e e e e} \}$ and $\{{\rm e e e e, E E E E E E}, {\it l l l l
}\}$.}
\end{figure}

Each causal class has univocally associated its {\it dual} causal
class \cite{199}. For example the Coll causal classes are dual
each other. The same occurs for the causal classes of the
Eddington-Finkelstein coordinates at $r>2m$ and $r<2m$. When a
causal class and its dual are equal, the class is said {\it
self-dual} (as the class of the Eddington-Finkelstein coordinates
at $r=2m$). Orthonormal frames and real null frames belong,
respectively, to the self-dual classes $\{{\rm t e e e, T T T E E
E}, {\it e e e e} \}$ and $\{{\rm l l e e, T L L L L E}, {\it l l
e e} \}$. There exist eleven self-dual causal classes which are
easily located looking at the diagonal region of the 199-Table.
They are:

\begin{displaymath}
\begin{array}{cc}
\{{\rm e\, e\, e \, e, \, T \, T \, T \, E \, E \, E,} \, {\it e
\, e \, e \, e} \} \quad & \quad \{{\rm l\, e\, e \, e, \, T \, T
\, T \, E \, E \, E,} \, {\it l \, e \, e \, e} \}\\

\{{\rm e\, e\, e \, e, \, T \, T \, L \, L \, E \, E,} \, {\it e
\, e \, e \, e} \} \quad & \quad \{{\rm l\, e\, e \, e, \, T \, T
\, L \, L \, E \, E,} \, {\it l \, e \, e \, e} \} \\

\{{\rm e\, e\, e \, e, \, T \, L \, L \, L \, L \, E,} \, {\it e
\, e \, e \, e} \} \quad & \quad \{{\rm l\, e\, e \, e, \, T \, T
\, L \, E \, L \, E,} \, {\it l \, e \, e \, e} \} \\

\{{\rm e\, e\, e \, e, \, L \, L \, L \, L \, L \, L,} \, {\it e
\, e \, e \, e} \} \quad & \quad  \\
\\
\{{\rm e\, l\, e \, e, \, T \, T \, E \, L \, L \, E,} \, {\it l
\, e \, e \, e} \} \quad & \quad \{{\rm t\, e\, e \, e, \, T \, T
\, T \, E \, E \, E,} \,\, {\it t \, e \, e \, e} \} \\

\{{\rm e\, l\, e \, e, \, T \, L \, L \, L \, L \, E,} \, {\it l
\, e \, e \, e} \} \quad & \quad \{{\rm l\, l\; e \,
e, \, \, T \, L \, L \, L \, L \, E,} \, \,{\it l \, l \; e \, e}\,\}\\
\end{array}
\end{displaymath}

To gain more comprehension of the role that coordinates play in
Relativity, the 199 causal classification would be investigated
through and through. The causal classes may be associated to
admissible {\it cuts} of the space-time others than the well-known
three-space $\oplus$ one-time usual in the at present evolution
conception of physics. Other cuts (among the other 198 possible
ones) may help us to better understand other aspects of the
space-time, and even to wake up our interest for other variations
of physical fields than the time-like ones associated to the
evolution formalism. A lot of work still remains to be done in
this direction.
\begin{theacknowledgments}
 This work has been supported by the Spanish Ministerio de Educación y
Ciencia, MEC-FEDER project AYA2003-08739-C02-02.

\end{theacknowledgments}

\end{document}